\documentclass[%
 reprint,
 superscriptaddress,
 amsmath,amssymb,
 aps,
]{revtex4-2}

\usepackage{graphicx}
\usepackage{dcolumn}
\usepackage{bm}
\usepackage{hyperref}
\usepackage[mathlines]{lineno}

\usepackage{lipsum}
\usepackage[dvipsnames]{xcolor}

\begin{document}

\preprint{APS/123-QED}

\title{
On the study of new proxies for 
second order cumulants of conserved charges \texorpdfstring{\\}{}
in heavy-ion collisions with EPOS4 
}

\author{Johannes Jahan}
\email{jahan.johannes@gmail.com}
\affiliation{%
Department of Physics - University of Houston, Houston, TX 77204, USA
}
\affiliation{
SUBATECH, University of Nantes - IN2P3/CNRS - IMT Atlantique, Nantes, France
}%

\author{Claudia Ratti}
\affiliation{%
Department of Physics - University of Houston, Houston, TX 77204, USA
}%

\author{Maria Stefaniak}
\affiliation{
Department of Physics - The Ohio State University, Columbus, OH 43210, USA
}%
\affiliation{
GSI Helmholtz Centre for Heavy Ion Research, 64291 Darmstadt, Germany
}%

\author{Klaus Werner}
\affiliation{
SUBATECH, University of Nantes - IN2P3/CNRS - IMT Atlantique, Nantes, France
}%

\date{\today}

\begin{abstract}
Proxies for cumulants of baryon number $B$, electric charge $Q$ and strangeness $S$ are usually measured in heavy-ion collisions via moments of net-number distribution of given hadronic species. 
Since these cumulants of conserved charges are expected to be sensitive to the existence of a critical point in the phase diagram of nuclear matter, it is crucial to ensure that the proxies used as substitutes are as close to them as possible.
Hence, we use the EPOS4 framework to generate Au+Au collisions at several collision energies of the RHIC Beam Energy Scan. We compute $2^\text{nd}$ order net-cumulants of $\pi$, $K$ and $p$, for which experimental data has been published, as well as the corresponding conserved charge cumulants. We then compare them with proxies, defined in previous lattice QCD and Hadron Resonance Gas model studies, which are shown to reproduce more accurately their associated conserved charge cumulants. 
We investigate the impact of hadronic re-scatterings occurring in the late evolution of the system on these quantities, as well as the amount of signal actually originating from the bulk medium which endures a phase transition.
\end{abstract}

\maketitle


\section{\label{sec:Intro}Introduction}

For several decades now, heavy-ion collisions (HICs) at relativistic energies have been used to study the properties of nuclear matter, and to map the phase diagram of Quantum Chromodynamics (QCD). Since the creation of a deconfined state of quarks and gluons at very high temperature, the Quark-Gluon Plasma (QGP), has been brought to light by several experiments at the Relativistic Heavy-Ion Collider (RHIC) \cite{BRAHMS:2004adc, PHOBOS:2004zne, STAR:2005gfr, PHENIX:2004vcz}, understanding the nature of the transition from a hadronic phase to the QGP has been one of the main focuses in the field. 
On the theoretical side, first-principle calculations from lattice QCD simulations predict a crossover at low baryon chemical potential $\mu_B$ \cite{Aoki:2006we,HotQCD:2018pds, Borsanyi:2020fev}, which seems confirmed by experimental results from RHIC and the Large Hadron Collider (LHC) \cite{Niida:2021wut}. Several approaches predict a $1^\text{st}$ order phase transition at high $\mu_B$, which would imply the existence of a critical point on the QCD phase diagram \cite{Fischer:2018sdj, Fuseau:2019zld, Fu:2019hdw, Hippert:2023bel}. 
The RHIC Beam Energy Scan (BES) program has thus been set up to scan a large region of the phase diagram at finite $\mu_B$, by colliding $^{197}_{\;\,79}$Au nuclei in a wide range of center-of-mass energies. They have collected data at collision energies from $\sqrt{s_{NN}} = 200$ GeV down to 7.7 GeV in collider mode through phases I and II \cite{Bzdak:2019pkr,Tlusty:2018rif}, and even down to $\sqrt{s_{NN}} = 3.0$ GeV through the fixed target program \cite{Meehan:2016qon}. Indeed, one of the main goals of this program is to search for the existence of a $1^\text{st}$ order phase transition, and thus to asses the location of the hypothetical critical point. 
\\

To probe the existence of a critical point in the QCD phase diagram, susceptibilities of conserved charges such as baryon number $B$, electric charge $Q$ and strangeness $S$ are the theoretical quantities of choice, as they diverge in the vicinity of the critical point. They can be calculated as derivatives of the pressure $p$ with respect to the chemical potential of the corresponding conserved charge \cite{Bellwied:2015lba}:
\begin{equation}
    \chi^{BQS}_{ijk} = 
    \frac{\partial^{i+j+k}(p/T^4)}{(\partial\hat{\mu}_B)^i (\partial\hat{\mu}_Q)^j (\partial\hat{\mu}_S)^k} \; ,
    \label{eq:susceptibilities}
\end{equation}
where $\hat{\mu}_i = \mu_i / T$. Because the pressure is directly related to the partition function, they quantify how much the latter is modified under variations of the different chemical potentials, and are thus sensitive to radical changes in the state of nuclear matter, \textit{e.g.} the effect of a first order phase transition or a critical point. 
One can also relate the susceptibilities to net-cumulants of conserved charges. For the case of second order cumulants, that will be at the center of our study, these relationships are: 
\begin{gather}
    \chi^{XY}_{11} 
    = \frac{\langle N_X N_Y \rangle - \langle N_X \rangle\langle N_Y \rangle}{VT^3} 
    = \frac{\sigma^{XY}_{11}}{VT^3}  \; ,
    \\
    \chi^{X}_{2} 
    = \frac{\langle {N_X}^2 \rangle - \langle N_X \rangle^2}{VT^3} 
    = \frac{\sigma^{2}_{X}}{VT^3}  \; ,
    \label{eq:(co)variances}
\end{gather}
where $\sigma^{X}_{2}$ and $\sigma^{XY}_{11}$ are respectively variance and covariance of the net-numbers of conserved charges $N_{X/Y}$ \cite{Luo:2017faz}. In the context of HICs, the (co)variances of net-hadron multiplicities can be used as probes for conserved charges \cite{Stephanov:2008qz}. 
So far, because they are the most abundant and easiest species to identify in experiments, net-proton cumulants have been used as proxies for $B$ fluctuations \cite{STAR:2021iop}, net-kaon cumulants have been used as proxies for $S$ fluctuations \cite{STAR:2017tfy}, and net-multiplicity cumulants of combined identified hadrons have been used as proxies for $Q$ fluctuations \cite{STAR:2019ans}.
By convention, the net-numbers $N_b$ of any baryon specie $b$ are defined as the difference between number of baryons and number of anti-baryons $n_b - n_{\overline{b}}$, while for meson species $m$ the net-number $N_m$ is usually defined as the difference between positively charged mesons and negatively charged ones $n_{m^+} - n_{m^-}$. 
\\

However, using HICs to probe the phase diagram of nuclear matter comes with a lot of limitations. While on the theoretical side, most predictions are obtained from grand-canonical ensemble calculations, assuming thermal equilibrium and infinite volume, the system created in heavy-ion collisions is very short-lived, finite, small-sized and experiences rapid expansion \cite{Bluhm:2020mpc}. 
Moreover, the volume and temperature are not accessible through measurements, and are changing during the evolution of a single event. It has been thus proposed to use ratios of net-cumulants to compare with theoretical predictions of susceptibilities, in order to get rid of the volume and temperature dependence at leading order \cite{Koch:2005vg}.
Finally, using particles as experimental probes shows severe limitations, first of all because not all hadrons can be experimentally detected. For example, neutral hadrons can carry e.g. baryon number or strangeness (like neutrons and $\Lambda$ baryons). Besides, measuring conserved charges through hadron distributions includes additional fluctuations, since particle production is a quantum process, thus probabilistic in essence.

It is hence important to find hadronic proxies that reflect conserved charge fluctuations as accurately as possible through the cumulants of their net-multiplicity distributions. In this paper, we want to address this question for $2^\text{nd}$ order cumulants of conserved charges, by comparing proxies measured by STAR in \cite{STAR:2019ans} against new proxies inspired by \cite{Bellwied:2019pxh} and with their corresponding conserved charge quantities, measured in Au+Au simulations in the RHIC beam energy range. We will employ the EPOS4 event generator \cite{Werner:2023zvo}, and take advantage of its modular design to compare the several observables of interests as measured through the final-state particle distribution with the same quantities at hadronisation time, to quantify the impact of the hadronic re-scatterings on them. Finally, as EPOS4 relies on a separate treatment of the bulk matter and high-$p_T$ particles to model the presence of a fluid in the system, as an imitation of the QGP, we will check how much of the signal of these cumulant proxies actually originates from the deconfined medium.
\\ 

In Section \ref{sec:Observables}, we will start by describing the different observables we will use in our work, before to introduce EPOS4 in Section \ref{sec:EPOS}, the event generator used to produce the simulations of Au+Au collisions on which rely our study. Then, after having discussed the details of our analysis and the simulations in Section \ref{sec:Method}, we will present our results and their interpretation them in Section \ref{sec:Results}, to finally summarise the conclusions of our study and mention potential outlooks in Section \ref{sec:Conclusion}.

\section{\label{sec:Observables}Proxies of conserved charge cumulants}

As mentioned in Section \ref{sec:Intro}, one way to probe fluctuations of conserved charges in HICs is to measure fluctuations in the net-distribution of hadronic species used as proxies. 
However, choosing appropriate proxies is essential to ensure that the observables we study reflect properly the corresponding conserved charge fluctuations. This is the motivation of our study. In this section, we present the commonly used experimental proxies at first, before introducing the new ones we propose because of their more accurate description of conserved charges in the context of HICs. Our new proposed proxies, taken in part from a previous work \cite{Bellwied:2019pxh}, will then be tested against the ones commonly measured in experiment, as well as the corresponding cumulants of conserved charges, through realistic simulations of HICs collisions generated with EPOS4.  

\subsection{\label{subsec:STAR_proxies} STAR proxies}

The experimental results on which this study is based are $2^\text{nd}$ order diagonal and off-diagonal cumulants (also referred to as variances and covariances) from net-$\pi$, net-$K$ and net-$p$ distribution, published by the STAR collaboration in \cite{STAR:2019ans}. They present, for Au+Au collisions at $\sqrt{s_{NN}} = 7.7$, 11.5, 14.5, 19.6, 27, 39, 62.4 and 200 GeV, the different (co)variances:
\begin{equation}
    \label{eq:STAR_covariances}
    \begin{pmatrix}
    \sigma_{\pi}^{2} & \sigma_{\pi p}^{} & \sigma_{\pi K}^{} \\
    \\
    " & \sigma_{p}^{2} & \sigma_{pK}^{} \\
    \\
    " & " & \sigma_{K}^{2}
    \end{pmatrix}{} \; ,
\end{equation}
within a pseudorapidity window $|\eta| < 0.5$ and for particles with $0.4 < p_T < 1.6$ GeV. Results are notably shown as a function of the centrality class for $0-80\%$ centrality (through the mean number of participant nucleons $\langle N_{part} \rangle$) for each collision energy, corrected for centrality bin-width effects (CBWE) \cite{Sahoo:2012wn}. These (co)variances are then used to build proxy ratios for net-charge cumulant ratios, in order to get rid of the explicit dependence on $T$ and $V$, which are not experimentally accessible event-by-event. The proxy ratios defined by STAR are the following (in parenthesis we indicate the corresponding ratio of conserved charge fluctuations they are meant to represent): 
\begin{gather}
    C_{Qp} = \frac{\sigma^{11}_{Qp}}{\sigma^{2}_{p}} \left( \cong \frac{\chi^{QB}_{11}}{\chi^{B}_{2}} \right)
    \;\; , \;\;
    C_{QK} = \frac{\sigma^{11}_{QK}}{\sigma^{2}_{K}} \left( \cong \frac{\chi^{QS}_{11}}{\chi^{S}_{2}} \right)
    \; , \nonumber \\
    C_{pK} = \frac{\sigma^{11}_{pK}}{\sigma^{2}_{K}} \left( \cong \frac{\chi^{BS}_{11}}{\chi^{S}_{2}} \right)
    \; .
    \label{eq:ratios_STAR}
\end{gather}
These proxy ratios are displayed as functions of $\langle N_{part} \rangle$ too, but also as functions of collision energy for most central ($0-5\%$) and most peripheral ($70-80\%$) collisions. These last results are of particular importance as they are expected to hint at the existence of the critical point, since decreasing the collision energy increases the baryonic chemical potential of the system created in the collision. Note that any proxy involving $Q$ is calculated through measured net-numbers of identified charged $\pi$, $K$ and $p/\overline{p}$, and not directly from unidentified charged particle tracks. This is done to ensure a proper estimation of detection efficiency correction, as pointed out in \cite{Vovchenko:2021xcs,Chatterjee:2021hcr}.

\begin{figure}[]
    \centering
    \includegraphics[width=0.8\linewidth]{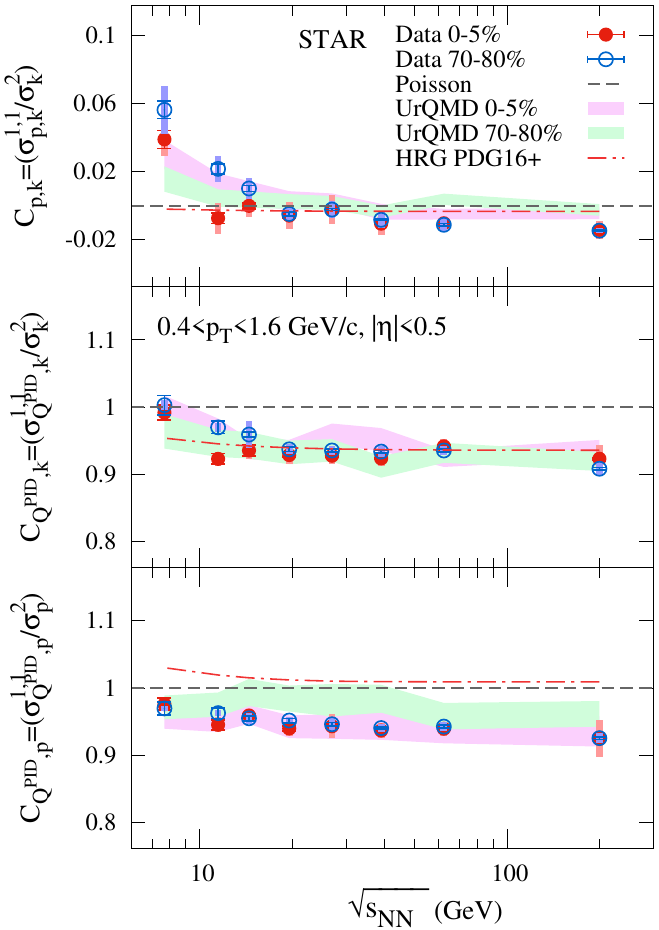}
    \caption{Energy dependence of the cumulants ratios $C_{pK}$, $C_{QK}$ for central (0-5\%) and peripheral (70-80\%) Au-Au collisions, within $|\eta| < 0.5$ and $0.4 < p_T < 1.6$ GeV. Results from STAR have systematic errors indicated by the boxes and statistical ones by bars. 
    Poisson baseline, UrQMD simulations results and calculations using HRG model including the PDG16+ \cite{Alba:2017mqu} particles set are displayed for comparison \cite{STAR:2019ans}.}
    \label{fig:STAR_proxy_ratios}
\end{figure}

One interesting observation from the results shown in Fig. \ref{fig:STAR_proxy_ratios} is the change of sign of the $C_{pK}$ ratio. While it shows no centrality dependence for $\sqrt{s_{NN}} \geq 19.6$ GeV, it changes sign at lower collision energies, where it also starts to show a centrality dependence. According to \cite{Koch:2005vg}, this could be due to a modification of the dynamics in the system, hinting for a change in the phase from which the studied correlations arise. This motivates even more the need to provide accurate proxies for conserved charge fluctuations, as this change of dynamics has also been shown through the observed disappearance of partonic collectivity in Au+Au collisions at $\sqrt{s_{NN}} = 3$ GeV \cite{STAR:2021yiu}.
\\

In this paper, we will systematically compare our results with the data from \cite{STAR:2019ans}, primarily as a benchmark to see if EPOS4 simulations are able to reproduce experimental results of net-multiplicity (co)variances for light hadrons. This will complement all the results from EPOS4 published so far, in particular the ones on particle production and flow for BES energies in \cite{Werner:2024ntd, Werner:2023mod}. Moreover, it is also interesting to compare those data with EPOS4 simulations and look for any significant deviation, because the latter can be considered as another non-critical baseline, as will be explained in Sec. \ref{sec:EPOS}.
\\

\subsection{\label{subsec:New_proxies} New improved proxies}

The work presented here has also been motivated by a theoretical study of proxies for fluctuations of $B$, $Q$ and $S$ \cite{Bellwied:2019pxh}. In this paper, the HRG model has been used to quantify how much every different hadronic species experimentally measured contributes to the lattice QCD susceptibilities, in the region where both approaches overlap. 
As HRG relies on a description of QCD matter in terms of a gas of non-interacting hadrons and resonances, it is by construction quite straightforward to express susceptibilities in terms of a sum of hadronic contributions:
\begin{multline}
    \chi^{BQS}_{ijk}(T,\Vec{\hat{\mu}}) = 
    \sum_R \sum_{h\,\in\,\text{stable}} (P_{R\to h})^l (B_h)^i (Q_h)^j (S_h)^k \\
    \times \frac{\partial^l p_h/T^4}{(\partial\hat{\mu}_B)^i (\partial\hat{\mu}_Q)^j (\partial\hat{\mu}_S)^k} 
    \; .
    \label{eq:chi_HRG}
\end{multline}
In this equation, the $l = i + j + k$ (with $l=2$ for our study). The $R$ index refers to resonances and $h$ to the stable hadrons these resonances can decay into. $P_{R\to h}$ represents the branching ratio for $R$ decaying into $h$, and contains implicitly an additional probabilistic fluctuation term arising from resonance decays \cite{Begun:2006jf}. $B_h$, $Q_h$ and $S_h$ are the respective baryon number, electric charge and strangeness of hadron specie $h$, and $p_h$ the pressure due to hadron specie $h$.
Thanks to the statistical nature of the HRG model, one can easily integrate the particle distributions over only a fraction of the complete phase space, in order to reproduce the finite coverage of detectors in experiments. 
\\

\begin{figure}[!h]
    \centering
    \includegraphics[width=0.9\linewidth]{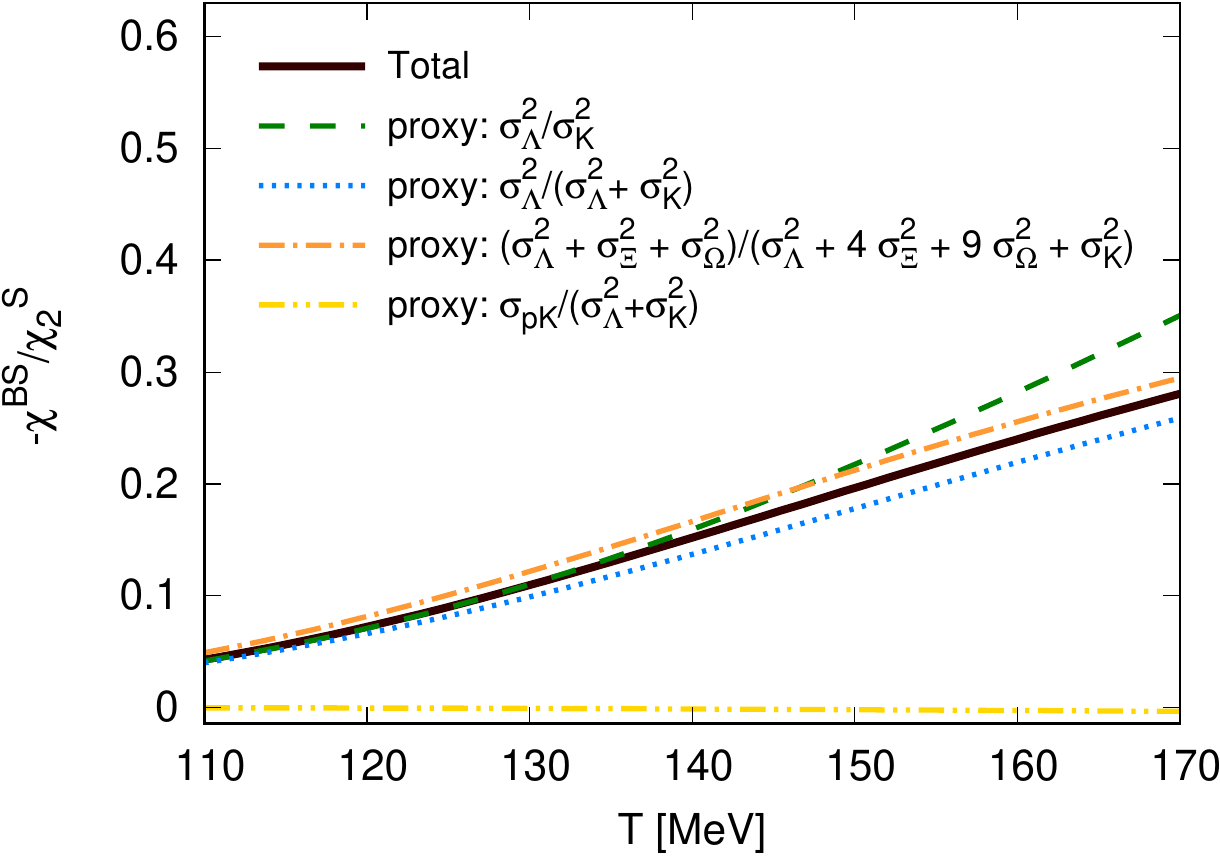}
    \caption{Comparison of several proxy ratios for $BS$ correlations obtained via HRG model calculations. Both are plotted as a function of $T$, with their related ratio of exact susceptibilities calculated with lattice QCD
    \cite{Bellwied:2019pxh}.}
    \label{fig:C_BS_breakdown}
\end{figure}

By doing so, the authors of \cite{Bellwied:2019pxh} have been able to construct new proxies based on the variance of net-distributions of some hadron species, that have been shown to reproduce susceptibility ratios very accurately. Some of these proxy ratios, denoted here as $\Tilde{C}_{QS}$ and $\Tilde{C}_{BS}(\Tilde{C}'_{BS})$ and respectively probing the same quantities as the STAR proxies ${C}_{QK}$ and ${C}_{pK}$ from eq. \eqref{eq:ratios_STAR}, are defined as:
\begin{align}
    & \hspace{0.5cm}
    \Tilde{C}_{QS} 
    = \frac{1}{2}. \frac{\sigma^2_K}{\sigma^2_K + \sigma^2_\Lambda} 
    \left( \cong \frac{\chi^{QS}_{11}}{\chi^{S}_{2}} \right) \; ,
    \label{eq:C_QS_HRG} \\
    \Tilde{C}_{BS} 
    & = \frac{\sigma^2_\Lambda}{\sigma^2_K + \sigma^2_\Lambda}
    \label{eq:C_BS_HRG}
    \hspace{0.3cm} / \\
    & \Tilde{C}'_{BS} 
    = \frac{\sigma^2_\Lambda + 2\sigma^2_\Xi + 3\sigma^2_\Omega}{ \sigma^2_\Lambda + 4\sigma^2_\Xi + 9\sigma^2_\Omega + \sigma^2_K}
    \left( \cong \frac{-\chi^{BS}_{11}}{\chi^{S}_{2}} \right) \; .    \label{eq:C_BS_HRGbis}
\end{align}
One can observe that these proxy ratios are only based on single hadron specie variances, and do not involve any covariance between different species. This is because cross-correlators between different conserved charges, $\chi^{XY}_{11}$, receive most of their contribution from the variances $\sigma^2_{h_1}$ of single species $h_1$ carrying both charges $X$ and $Y$. Only a very small fraction of $\chi^{XY}_{11}$ actually comes from covariances $\sigma_{h_1 h_2}$ between two different species $h_1$ and $h_2$, each of them carrying one of the two conserved charges $X$ and $Y$. 
Note also that two ratios, $\Tilde{C}_{BS}$ and $\Tilde{C}'_{BS}$, have been proposed as proxies for $\chi_{11}^{BS}/\chi_2^S$. Both of them are built in the same way, $\Tilde{C}'_{BS}$ being an extension of $\Tilde{C}_{BS}$ that includes in addition the contribution of multi-strange baryons. Even though this extra contribution in $\Tilde{C}'_{BS}$ (represented by the dashed-dotted orange line on Fig. \ref{fig:C_BS_breakdown}) does not improve significantly the situation compared to $\Tilde{C}_{BS}$ (shown by the dotted blue line), we will still compute both of them to see how they compare in event generator simulations.
\\

\begin{figure}
    \centering
    \includegraphics[width=0.9\linewidth]{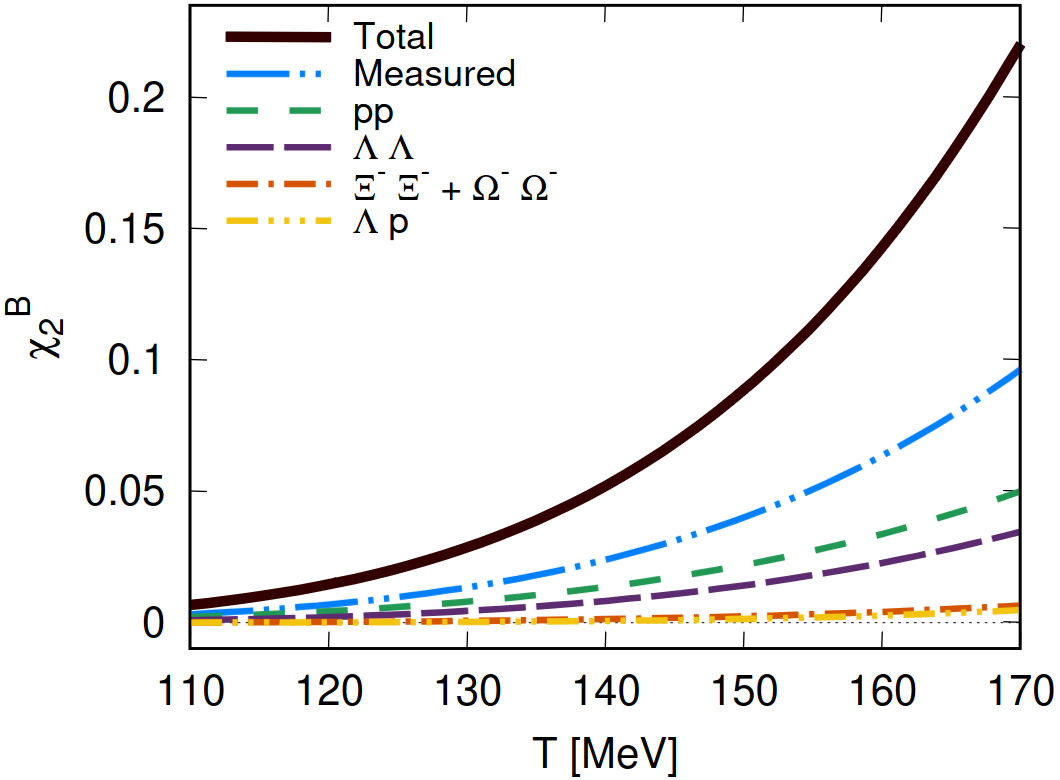}
    \caption{Breakdown of the different final state hadronic contributions to the diagonal correlators of $\chi^B_2$. The total contribution and the measured part are shown as solid black and dashed-dotted blue lines respectively. The main single contributions from measured hadronic observables are shown with different colored dashed and dashed-dotted lines \cite{Bellwied:2019pxh}.}
    \label{fig:chis_breakdown}
\end{figure}

However, no proxy for $\chi^{QB}_{11}/\chi^B_2$ has been proposed in \cite{Bellwied:2019pxh}, because authors claim that isospin randomization of the nucleons, due to reactions of the type: \cite{Kitazawa:2011wh,Kitazawa:2012at}
\begin{equation*}
    p + \pi^{0/-} \leftrightarrow \Delta^{+/0} \leftrightarrow n + \pi^{+/0}
    \label{eq:isospin_rando}
\end{equation*}
happening in the hadronic phase to both baryons and anti-baryons,' and occurring for several cycles, prevents from building a good proxy for this ratio. Since STAR measured this quantity through the proxy $C_{Qp}$, introduced in Eq. \eqref{eq:ratios_STAR}, we still want to propose here a new proxy. 
We start with the numerator $\chi^{QB}_{11}(=\chi^{BQ}_{11})$, which cannot be probed by $\sigma^2_p$ alone since it is affected by isospin randomization, but instead by  $\sigma^{}_{p\pi} + \sigma^2_p$, which is not. The use of this quantity was already attempted in \cite{Bellwied:2019pxh}, when trying to construct a proxy ratio for $\chi^{QB}_{11}/\chi^Q_2$. 
Next, we need a proxy for the denominator $\chi^B_2$. Because isospin randomization leads to equilibrate the amount of protons and neutrons in the system, one can probe the nucleon variance using the proton one through $\sigma^2_{\text{Net-}N} = 2 \times \sigma^2_p$ \cite{Bellwied:2019pxh}. We propose to use the net-$\Lambda$ variance in addition, as it is the second major contributor to $\chi^B_2$ as shown in Fig. \ref{fig:chis_breakdown}, it is also used in the other proxy ratios \eqref{eq:C_QS_HRG}, \eqref{eq:C_BS_HRG} and \eqref{eq:C_BS_HRGbis}, and was already measured experimentally \cite{STAR:2020ddh}.
Therefore, we propose the proxy ratio:
\begin{equation}
    \Tilde{C}_{QB}
    = \frac{\sigma^{11}_{\pi p} + \sigma^{2}_{p}}{2\sigma^{2}_{p} + \sigma^{2}_{\Lambda}}
    \left( \cong \frac{\chi^{QB}_{11}}{\chi^{B}_{2}} \right) \; ,
    \label{eq:C_QB_HRG}
\end{equation}
to probe a similar quantity as the STAR proxy $C_{Qp}$ from eq. \eqref{eq:ratios_STAR}.
\\

\section{\label{sec:EPOS} EPOS4}

Event generators are essential computational tools in modern high-energy physics phenomenology, enabling to bridge the gap between theoretical models and experimental data \cite{Campbell:2022qmc}.
In this study, we employ EPOS4.0.0 \cite{Werner:2023zvo}, the latest version of the multi-purpose event generator, released publicly through a dedicated webpage \cite{EPOS4-website}.
EPOS4 is built to offer a realistic description of high-energy hadronic collisions from several GeV to several TeV per-nucleon center-of-mass energy. 
Testing the proxies for conserved charge fluctuations introduced in Section \ref{sec:Observables} aims thus at validating the study using lattice QCD data and HRG simulations, on which these proxies are based. In addition, achieving this study through realistic simulations of HICs with EPOS4 offers a more direct connection with the experimental scenario, and a meaningful comparison with experimental proxies already in use to probe fluctuations.
\\

In EPOS4, the primary interactions are modeled through a multiple-interaction approach based on the $S$-matrix theory including perturbative QCD calculations and saturation effects \cite{Drescher:2000ha}. In EPOS4, major improvements have been achieved by ensuring a consistent treatment of saturation and factorization, through the implementation of a dynamical saturation scale within this energy-conserving parallel scattering framework \cite{Werner:2023mod, Werner:2023zvo}. The treatment of perturbative calculations has also been updated, with a particular care for heavy flavors \cite{Werner:2023fne}.

\begin{figure}
    \centering
    \includegraphics[width=0.75\linewidth]{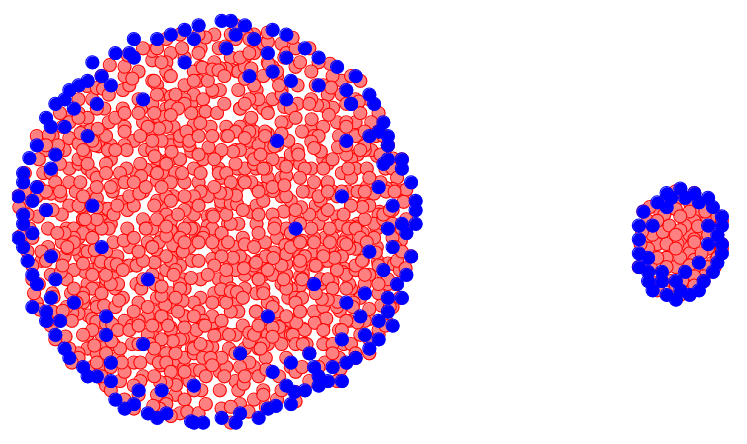}
    \caption{Sketch of the core-corona separation for a “big” and a “small” system. The dots are pre-hadrons in the transverse plane, with red referring to pre-hadrons from core, blue to pre-hadrons from corona \cite{Werner:2023jps}.}
    \label{fig:core-corona}
\end{figure}

The pre-hadrons formed through primary scatterings and remnants excitation are then split into 2 categories, following a so-called "core-corona" procedure, as shown in Fig. \ref{fig:core-corona}, based on their density and some energy loss considerations \cite{Werner:2007bf,Werner:2013tya}.
This procedure, applied at some initial time $\tau_0 = 0.4$ fm/c, has been discussed with many details in the context of EPOS4 in \cite{Werner:2023jps}.
The hot and dense bulk matter made from low transverse momentum pre-hadrons, the ``core", evolves following (3+1)D viscous hydrodynamics equations using vHLLE \cite{Karpenko:2013wva}, and employing a simple crossover equation of state (EoS) matched with lattice QCD data \cite{Werner:2013tya}. Once the system has expanded and cooled down, it will hadronize when reaching an energy density value $\epsilon_H = 0.57$ GeV/fm$^3$, based on a micro-canonical procedure \cite{Werner:2023jps}.
This is particularly important for simulating small systems, where few particles are produced, and hence local energy and charge conservation play an important role.
On the other hand, pre-hadrons with high-$p_T$ (sufficient to escape the core) form the ``corona", and may then still re-interact with the core hadrons through hadronic scatterings.

Hadronic re-scatterings between all formed hadrons (also referred to as hadronic cascades) are simulated using UrQMD as an after-burner \cite{Bleicher:1999xi, Petersen:2008dd}. It models elastic scatterings, string and resonance excitations as well as $2 \to n$ processes and strangeness exchange reactions based on measured reaction cross-sections, and includes 60 baryonic species + 40 mesonic states.
\\

For the work presented here, we take advantage of the modular construction of EPOS4, since one can access the distribution of particles at hadronization, just before hadrons are fed into UrQMD. This allows us to study the impact of hadronic cascades on fluctuation observables by comparing this distribution with the final-state particle one.
Moreover, the core-corona procedure applied in EPOS4 allows to differentiate particles originating from the bulk matter (the ``core" in the EPOS4 framework) and the ones originating from hard processes (``corona" hadrons). Hence, one can extract the proportion of the total fluctuation signal measured in the final-state that is actually coming from the core, which contains the potential critical fluctuations.
\\

We want to mention that, among the improvements brought to the EPOS framework in EPOS4, it is now possible to use an EoS including a 1$^\text{st}$ order phase transition and a critical point from the 3D Ising model universality class, such as e.g. the one published by the BEST collaboration \cite{Parotto:2018pwx} and its recent improvement \cite{Kahangirwe:2024cny}, to model the hydrodynamical evolution of the core \cite{Stefaniak:2022pxc}. 
However, we do not use it for the HIC simulations used in this work, but employ a simple crossover EoS as mentioned previously. The main reason is that our goal is to compare different proxy observables with their corresponding conserved charge quantities, and to study the impact of hadronic cascades on them. 
We do not aim at investigating the effect of criticality on these observables, since EPOS4 does not propose the adequate framework for such investigation. In fact, it employs an ``ordinary" hydrodynamical code that does not incorporate any out-of-equilibrium fluctuations nor charge dissipation relations, which are necessary to study the impact of critical behaviour on the evolution of the system created in HICs. Several works are currently ongoing towards the development of hydrodynamics framework that would properly take fluctuations into account  \cite{Stephanov:2017ghc, Pihan:2022xcl}, which might eventually be included in EPOS4 in the future.
Until then, EPOS4 can be considered as a non-critical baseline, especially when run with the crossover EoS employed here.
\\

\section{\label{sec:Method} Analysis details}

For all the results presented in Section \ref{sec:Results}, particle distributions have been evaluated using the same cuts employed by STAR in \cite{STAR:2019ans}, to allow direct comparison with their data.
We take into account all particle distributions within a pseudorapidity window $|\eta| < 0.5$ and for particles with $0.4 < p_T < 1.6$ GeV. 
To correct from feed-down, we chose to consider only particles that have a vertex of origin located at less than $1$ cm from the interaction vertex, instead of explicitly neglecting all secondary particles as EPOS4 would allow.
This is to stay as close to the experimental analysis as possible, where they correct from contamination of secondary particles by considering only tracks with a distance of closest approach (DCA) to the primary vertex $< 1$ cm.

\begin{table}[!h]
    \centering
    \begin{tabular}{|c||c|c|c|c|c|c|c|c|}
        \hline
        $\mathbf{\sqrt{s_{NN}}}$ \textbf{(GeV)}
        & \textbf{19.6} & \textbf{27} & \textbf{39} & \textbf{62.4} &  \textbf{200} \\
        \hline
        $\mathbf{N_{evts}}$ \textbf{} 
        & 2.5M & 2.5M & 2.5M & 2.0M & 1.5M \\
        \hline
    \end{tabular}
    \caption{Total number of individual Au+Au events per collision energy simulated with EPOS4.
    }
    \label{tab:event_stats}
\end{table}

In Table \ref{tab:event_stats}, one can find the number of events simulated with EPOS4 for each of the collision energies shown in this work.
Note that we will only show EPOS4 results from $\sqrt{s_{NN}} = 19.6$ GeV and above, although STAR data are available down to $\sqrt{s_{NN}} = 7.7$ GeV. 
We choose to do so since a previous study of EPOS4 results, in the energy range of the BES, showed that the model is not applicable anymore below such collision energy \cite{Werner:2024ntd}. 
In fact, a systematic study of $p_T$ spectra of many light and strange hadrons at several collision energies has shown that EPOS4 fails at reproducing, even qualitatively, the available experimental data below $\sqrt{s_{NN}} = 19.6$ GeV. This is expected, due to current limitations inherent to the approach employed to model the primary interactions. Thus, we will only discuss here results in a collision energy range which is physically relevant to the model.
All events used in this study have been generated using the core-corona separation procedure, the so-called x3FF crossover equation of state mentioned in Section \ref{sec:EPOS} with the vHLLE hydrodynamic model for the evolution of bulk matter, and neglecting any jet-fluid interaction.
\\

Another important aspect when studying cumulants of net-multiplicity distributions in HICs, is the definition of centrality classes, and how we correct cumulants from the CBWE \cite{Sahoo:2012wn}. For all results shown as functions of the number of participating nucleons $N_{part}$, we compute (co)variances directly as functions of this observable, since it is accessible through our simulations. 
However, we do not use the actual $N_{part}$ value from EPOS4 primary interaction model, but the Glauber estimated value based on the impact parameter. This way, we ensure the closest apple-to-apple comparison with the $N_{part}$ values estimated for each centrality class in the STAR analysis \cite{STAR:2019ans}. For these results, we divide our distributions into 20 bins of width $\Delta N_{part} = 20$, which broadly correspond to centrality classes of $\sim 5\%$.
The advantage is that, using directly $N_{part}$ to determine centrality, with such narrow bins in particular, reduces the volume fluctuations within a given centrality class compared with experimental centrality classes based on charged multiplicity. Moreover, a study of the CBWE has shown that second order cumulants should not be affected much by this effect, for centrality classes smaller than 10\% \cite{Sahoo:2012wn}. We have successfully tested this in EPOS4 simulations for collisions at $\sqrt{s_{NN}} = 200$ GeV, as shown in Appendix B of \cite{jahan:tel-04051507}. 
All results displayed as functions of collision energy $\sqrt{s_{NN}}$ for different centrality classes are obtained using the same centrality determination procedure. 
\\

The main goal of our study being to test the accuracy of the net-charge proxy cumulants introduced in Section \ref{sec:Observables}, we will compare them with the actual net-$B$, net-$Q$ and net-$S$ cumulants, which are computable from EPOS4 simulations because all particles are accessible. 
In order to make a relevant comparison with the proxy cumulants, we calculate the net-charge cumulants by considering the charges carried by all particles in the same phase space as the one used for the proxies and the STAR analysis, \textit{i.e.} within a pseudorapidity window $|\eta| < 0.5$ and for particles with $0.4 < p_T < 1.6$ GeV.
We apply the same DCA $<1$ cm criterion to select particles, in order to correct from feed-down contamination, and consider particles before weak decays, since the latter violates $S$ conservation. Technical details about the way we perform the calculation of net-charge cumulants in the EPOS4 simulated events are provided in Appendix C of \cite{jahan:tel-04051507}.

\section{\label{sec:Results}Results and discussion}

\begin{figure*}[!ht]
    \centering
    \includegraphics[width=1\linewidth]{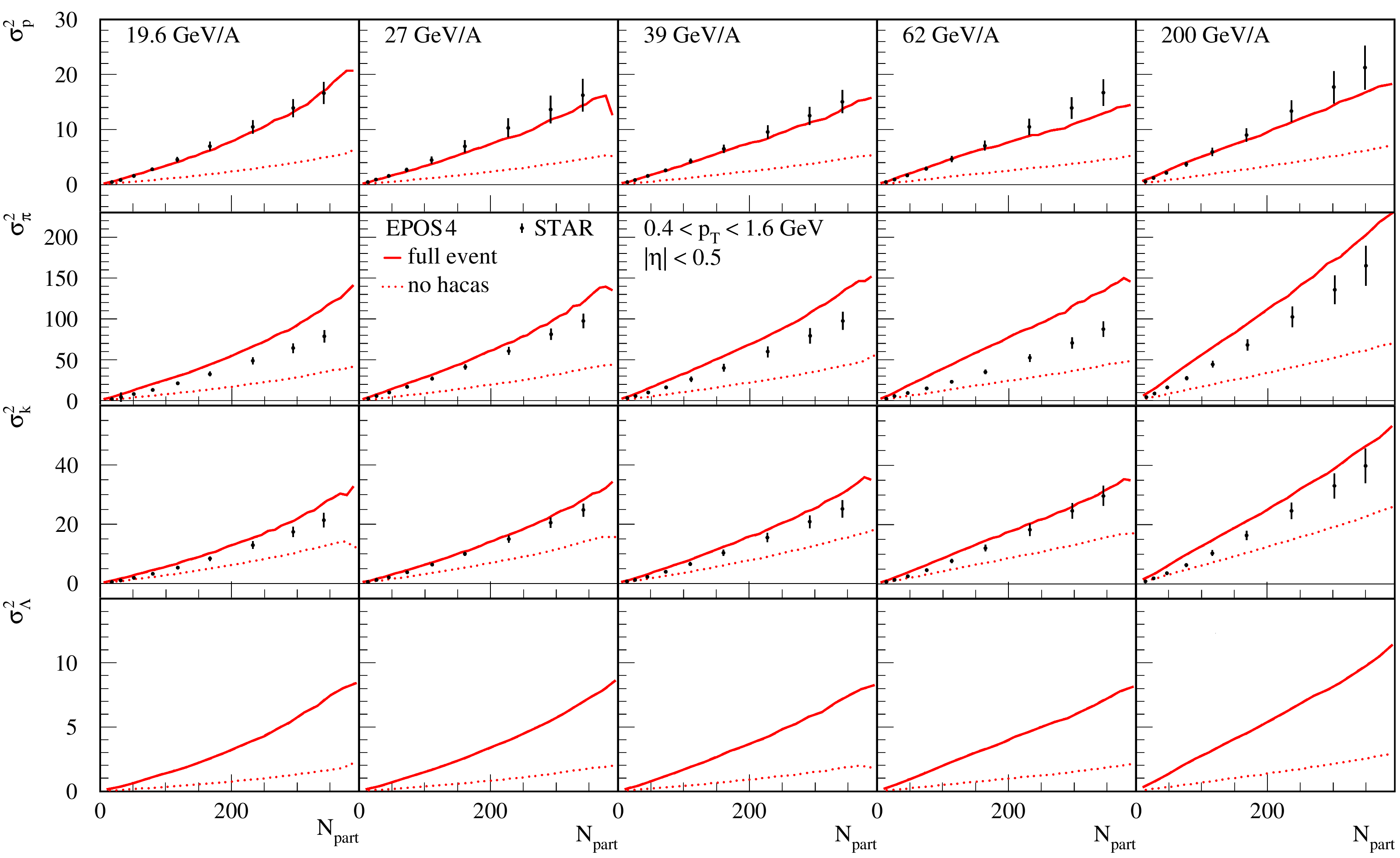}
    \caption{
    Net-$p$, net-$\pi$, net-$K$ and net-$\Lambda$ multiplicity variances vs. $N_{part}$ for different energies, obtained from EPOS4 simulations and compared with STAR results (dots) \cite{STAR:2019ans}. The results calculated from full events are represented by full lines, while results calculated before hadronic cascades are represented by dotted lines.
    }
    \label{fig:variances_nohacas_EPOS4}
\end{figure*}

In this section, we discuss the results of our study of the proxies for $2^\text{nd}$ order cumulants of conserved charges, obtained through the analyses described in Section \ref{sec:Method}. 
We will first show the (co)variances of different identified hadron species at several collision energies $\sqrt{s_{NN}}$, as functions of $N_{part}$, for final-state particle distributions and also particle distributions just after hadronization, compared with experimental data from STAR \cite{STAR:2019ans}. 
From there, we will build the different proxy ratios, and compare them with their corresponding conserved charge cumulant ratios, as well as with STAR data, still as functions of $N_{part}$ for different collision energies. 
In order to discuss the results of our study in a more summarized way, we will then show the different ratios as functions of $\sqrt{s_{NN}}$, for the most central collisions. 
At first, we will show results for both final-state particle distributions and particle distributions at hadronization, to see the impact of hadronic cascades on displayed quantities. 
Finally, we will show the proxy and corresponding conserved charge ratios for distributions of all particles at hadronization, compared with the same quantities but only from core particles, in order to highlight the contribution of bulk matter to the fluctuation signal measured in the phase space considered in this analysis. 
\\

More detailed results on particle production relevant for the system and collision energies studied in this paper, based on the same set of simulated events, can be found in \cite{Werner:2024ntd, Werner:2023mod}.

\subsection{\label{subsec:results_(co)variances}Variances and covariances of identified \\ hadron net-multiplicities}

In Fig. \ref{fig:variances_nohacas_EPOS4}, we show the centrality dependence (via the number of participants $N_{part}$) of variances of net-$\pi$, net-$K$, net-$p$ and net-$\Lambda$ multiplicities at collision energies $\sqrt{s_{NN}} = 19.6$, 27, 39, 62.4 and 200 GeV. There, one can see variances from final-state particle distributions (full lines) and from particle distributions at hadronization (dashed lines), both from EPOS4, compared with STAR results (dots) \cite{STAR:2019ans}.
We observe that EPOS4 reproduces qualitatively the expected (quasi-)linear increase with decreasing centrality (increasing $N_{part}$) of the different variances in both distributions. 
On a quantitative level, EPOS4 results for final-state distributions do  not match perfectly the data from STAR, although describing them quite well overall. 
This is however not of extreme importance in the context of this work, since we only aim at making a quantitative comparative study.
Another important aspect to notice in this Figure, is the difference between variances for particle distribution at hadronization and in the final-state. For all hadronic species shown, hadronic cascades increase the variance at all collision energies, by a factor unique to each species which is quite constant regardless of the energy considered.
They increase the magnitude of the signal by a factor $\sim2$ for $\sigma^2_K$ to $\sim3$ for $\sigma^2_\pi$ and $\sigma^2_p$, even up to a factor $\sim4$ for $\sigma^2_\Lambda$ where the effect is the most significant. 
\\

\begin{figure*}
    \centering
    \includegraphics[width=1\textwidth]{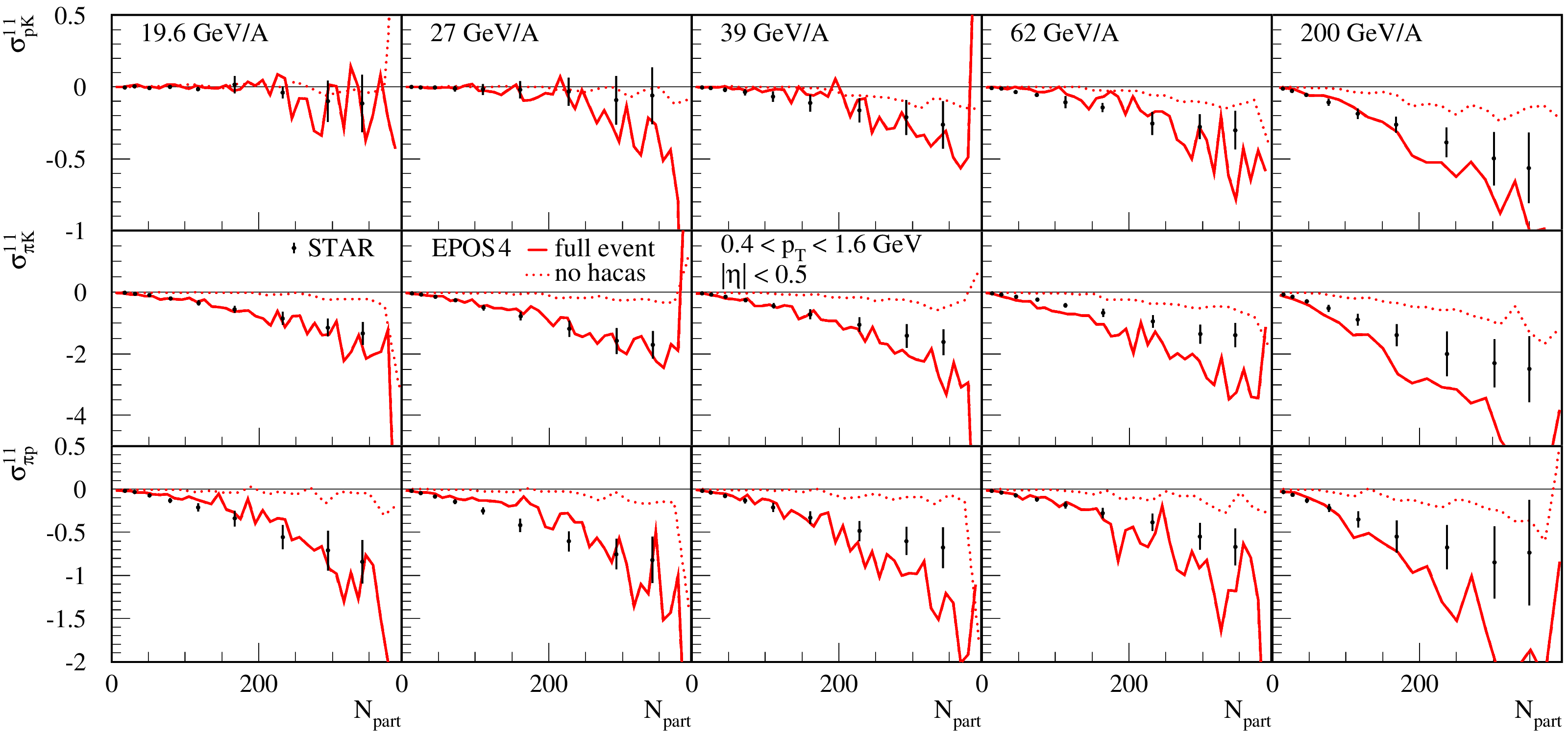}
    \caption{
    Covariances of net-$p$, net-$\pi$ and net-$K$ multiplicity distributions vs. $N_{part}$ for different energies, obtained from EPOS4 simulations and compared with STAR results (dots) 
    \cite{STAR:2019ans}. Results calculated from full events are represented by full lines, while results calculated before hadronic cascades are represented by dotted lines.
    }
    \label{fig:covariances_nohacas_EPOS4}
\end{figure*}

In Fig. \ref{fig:covariances_nohacas_EPOS4}, the same plots are shown for the different covariances $\sigma^{11}_{\pi p}$, $\sigma^{11}_{\pi K}$ and  $\sigma^{11}_{pK}$.
Despite the apparent wiggles observed in the results for covariances, caused by the limited statistics used for this analysis (see Table \ref{tab:event_stats}), one can conclude similarly to the case of variances that the results from EPOS4 final-state distributions reproduce the qualitative behavior of STAR data, and are reasonably close on a quantitative level.
We also note an increase of the absolute amplitude of the signal due to hadronic cascades, when comparing EPOS4 distributions at hadronization (again shown as a dashed line) with EPOS4 distributions of final-state hadrons (full line). 
\\

The observed increase of (co)variances can be explained by the fact that, during hadronic cascades, the numerous inelastic scatterings produce many more of the involved hadronic species, coupled to the successive decays of heavy resonances produced at hadronization. Hence, the successive decays and inelastic scatterings, producing simultaneously several of the hadronic species of interest, lead unavoidably to the amplification  of the correlations carried by (co)variances, blurring the actual magnitude of the thermal fluctuations of interest by increasing the signal.
These observations support the results of a study led with a hybrid version of UrQMD, which has been coupled to a fluid dynamical simulation of bulk matter \cite{Steinheimer:2016cir}. The authors of \cite{Steinheimer:2016cir} investigated the time-evolution of the correlation of distributions of hadron families carrying conserved charges, compared to the same distributions at hadronization time. The conclusion was that the final distribution of hadrons observed by experiments is not really related anymore to the one obtained after hadronization, which would contain the potential critical signal from the QCD phase transition.
\\

\subsection{Ratios of $B,Q,S$ cumulants and associated proxies}

\begin{figure*}
    \centering
    \includegraphics[width=1\textwidth]{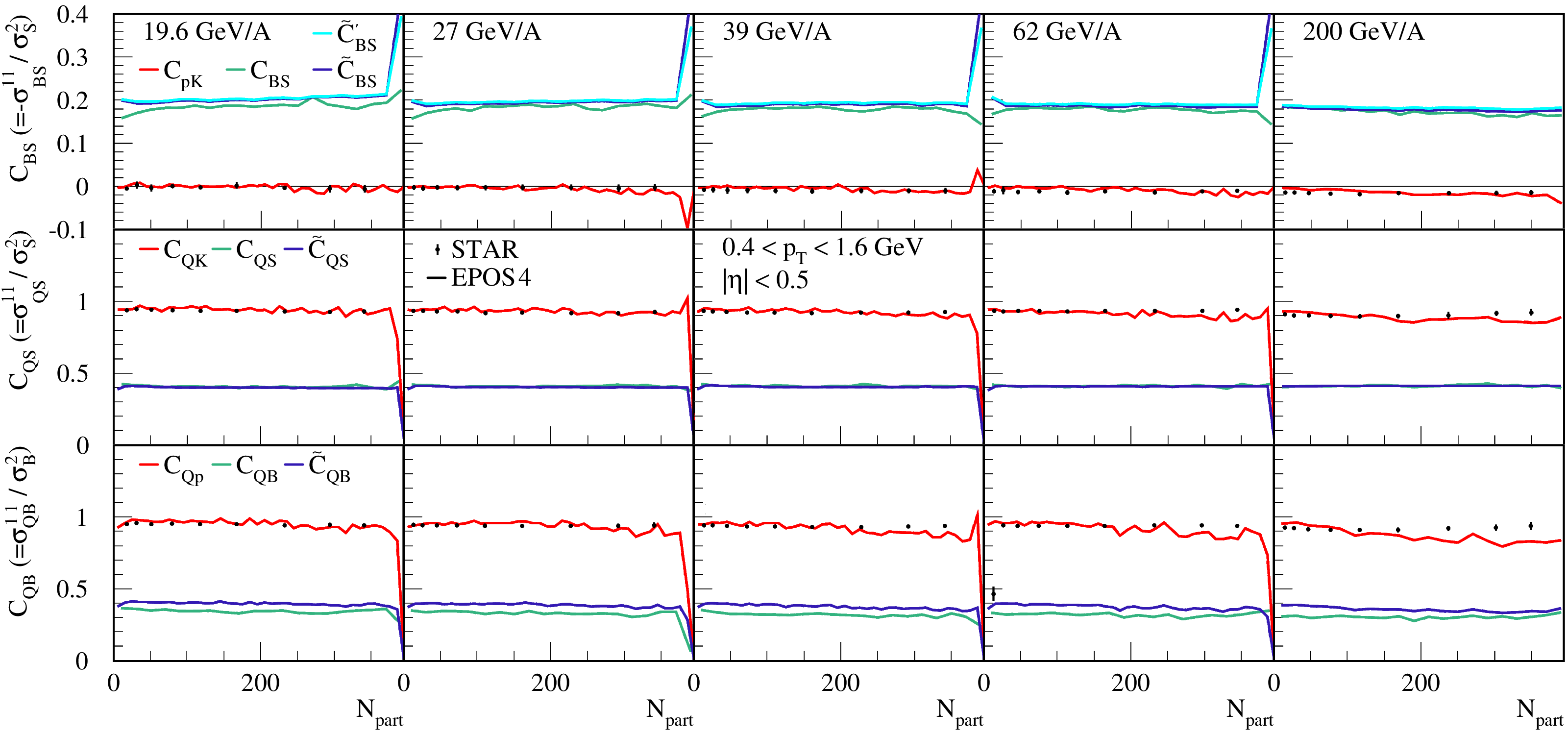}
    \caption{
    Ratios of exact $B$, $Q$ and $S$ conserved charge (co)variances and their proxies displayed as functions of $N_{part}$ for different energies, obtained from EPOS 3.451 simulations (lines) and compared with STAR data (dots) as a reference \cite{STAR:2019ans}.
    }
    \label{fig:ratios_Npart_proxies_EPOS4}
\end{figure*}

In Fig. \ref{fig:ratios_Npart_proxies_EPOS4}, we show the different ratios of $2^\text{nd}$ order conserved charge cumulants with their corresponding proxies, both from EPOS4 and STAR measurements from \cite{STAR:2019ans}, as functions of $N_{part}$ for different collision energies.
The red curves correspond to the STAR proxy ratios $C_{Qp}$, $C_{QK}$ and $C_{pK}$ measured in EPOS4 simulations (defined in Eq. \eqref{eq:ratios_STAR}), to be compared with STAR measurements of the same quantities shown as black dots.
For each of these 3 quantities, we also plot in dark blue the corresponding new proposed proxies $\Tilde{C}_{QB}$, $\Tilde{C}_{QS}$ and $\Tilde{C}_{BS}$ + $\Tilde{C}'_{BS}$ (in light blue), defined in Eqs. \eqref{eq:C_QB_HRG}, \eqref{eq:C_QS_HRG} and \eqref{eq:C_BS_HRG} + \eqref{eq:C_BS_HRGbis} respectively. Finally, we display the corresponding ratios of exact cumulants of conserved charges in green, as a reference. Note that we will use this same color code for all figures displaying cumulants ratios.
\\

Let us discuss first the comparison of STAR proxies measured in EPOS4 simulations, with the actual data from \cite{STAR:2019ans} for those quantities.
Although variances and covariances computed from EPOS4 simulations do not reproduce perfectly the STAR data, as discussed in the previous subsection, we notice in Fig. \ref{fig:ratios_Npart_proxies_EPOS4} that the proxy ratios $C_{Qp}$, $C_{QK}$ and $C_{pK}$ constructed from them do match STAR data point very well on a quantitative level.

Another noticeable feature is the apparent absence of centrality dependence, observed for all ratios displayed in Fig. \ref{fig:ratios_Npart_proxies_EPOS4} (apart from a slight decrease of the ratio $C_{Qp}$ with increasing $N_{part}$, that appears to be more significant with increasing collision energy). For this reason, we will compare from now on the different ratios of conserved charge cumulants with their corresponding proxies as a function of collision energy, for central collisions (0-5\% centrality class) only.

Finally, we want to address the case of the ratio $C_{BS}$ $(=\sigma^{11}_{BS}/\sigma^2_S)$ and its proxies. As an alternative to the STAR proxy $C_{pK}$, we have decided to consider two proxy ratios from \cite{Bellwied:2019pxh}, $\tilde{C}_{BS}$ and $\tilde{C}'_{BS}$. 
While the first one, defined in Eq. \eqref{eq:C_BS_HRG}, relies only on net-$\Lambda$ variance as a carrier of both strangeness and baryon charge, the second one defined in Eq. \eqref{eq:C_BS_HRGbis} also incorporates variances of net-$\Xi$ and net-$\Omega$.
However, despite the fact that $\Xi$ and $\Omega$ baryons carry respectively a strangeness content of $|S|=2$ and $|S|=3$, they are produced at a significantly lower rate compared to $\Lambda$ baryons in HICs, approximately $1$ and $2$ orders magnitude less, respectively. 
This low relative abundance makes their contribution to  the ${C}_{BS}$ proxy ratio almost insignificant in comparison with the net-$\Lambda$ variance. Looking at the top panels of Fig. \ref{fig:ratios_Npart_proxies_EPOS4}, showing $C_{BS}$ and related proxies, one can effectively see that $\tilde{C}_{BS}$ and $\tilde{C}'_{BS}$ results completely overlap with over the whole centrality range and all collision energies displayed. Thus, we choose to discard $\tilde{C}'_{BS}$ from the results from now on, and only focus on $\tilde{C}_{BS}$ as a new alternative proxy ratio to ${C}_{BS}$.
\\

\begin{figure}
    \centering
    \includegraphics[width=0.8\linewidth]{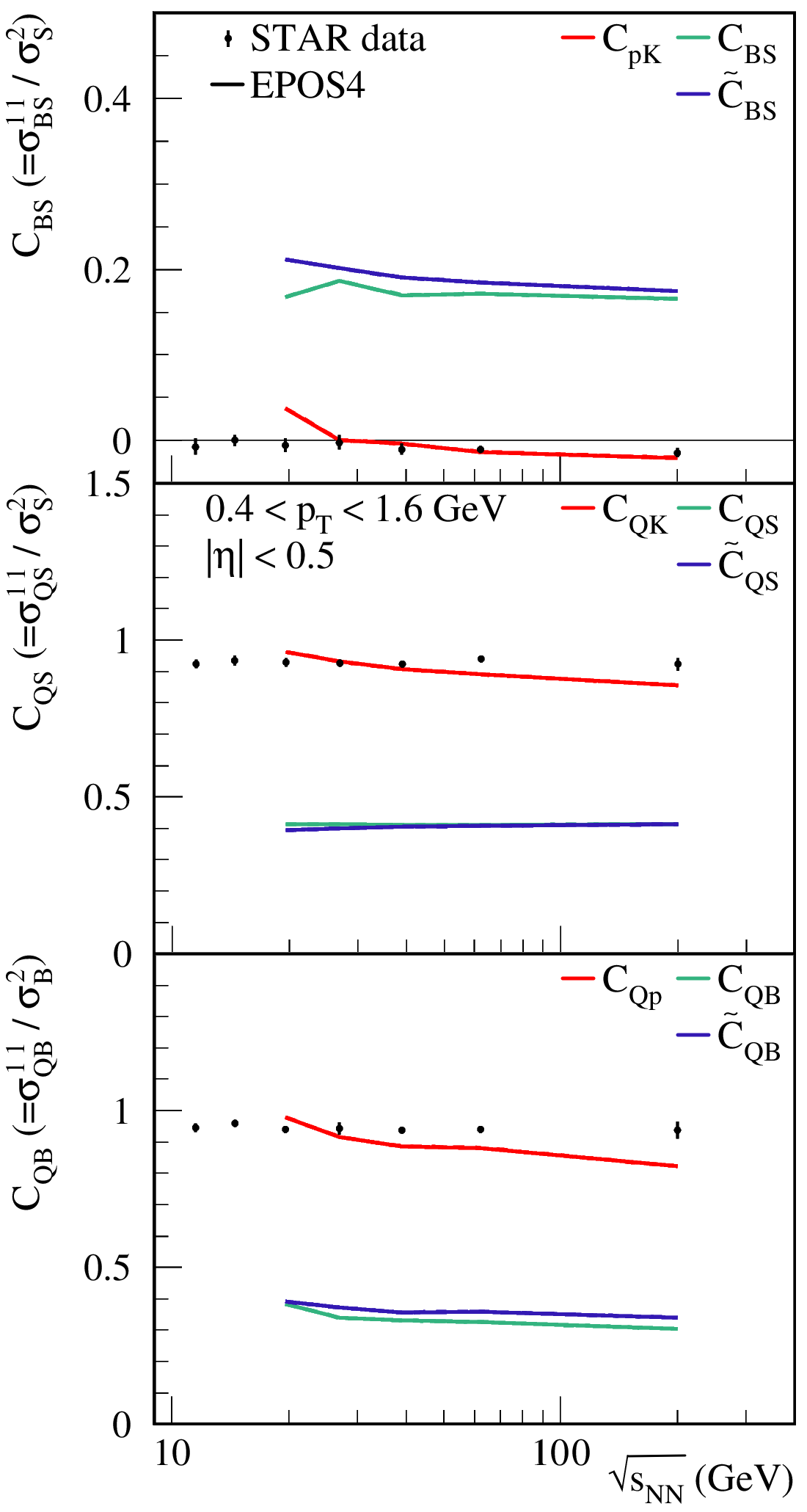}
    \caption{
    Ratios of exact $B$, $Q$ and $S$ conserved charge (co)variances and their proxies, displayed as functions of $\sqrt{s_{NN}}$ for central collisions (0-5\% centrality class). Results from EPOS4 simulations for final-state particle distributions (lines) are compared with STAR data (dots) as a reference \cite{STAR:2019ans}.
    }
    \label{fig:ratios_engy_proxies_EPOS4}
\end{figure}

Coming now to the first main results of our study, we show in Fig. \ref{fig:ratios_engy_proxies_EPOS4} the ratios $C_{BS}$, $C_{QS}$ and $C_{QB}$ compared with their respective proxy ratios and STAR results, as functions of collision energy $\sqrt{s_{NN}}$ for 0-5\% centrality class events. We observe that the STAR proxy ratios from Eq. \ref{eq:ratios_STAR} measured in EPOS4 simulations are very close to the STAR data (as already noticed previously), despite a slight energy dependence not present in the data. 
Overall, we note that none of the displayed ratios shows any clear collision energy dependence, for the range that is shown here.
However, the important result in these plots lies in the comparison between the ratios of conserved charge cumulants (in green), taken as a reference, and their proxies. STAR data are only shown as a benchmark, to see how well EPOS4 can reproduce experimental results. 
One can observe how remarkably well the new proposed proxies $\tilde{C}_{XY}$ (in blue) are able to reproduce quantitatively the conserved charge cumulant ratios, compared to the STAR proxies $C_{QpK}$ (in red) which show a clear deviation in magnitude.
These new proxies are thus shown to be better quantitative probes than the ones traditionally used by STAR, as well as other experiments to probe fluctuations through $2^\text{nd}$ order cumulants of conserved charges in HICs (see \cite{MUSES:2023hyz} for a review of these measurements).

\subsection{Impact of hadronic cascades}

\begin{figure}
    \centering
    \includegraphics[width=0.8\linewidth]{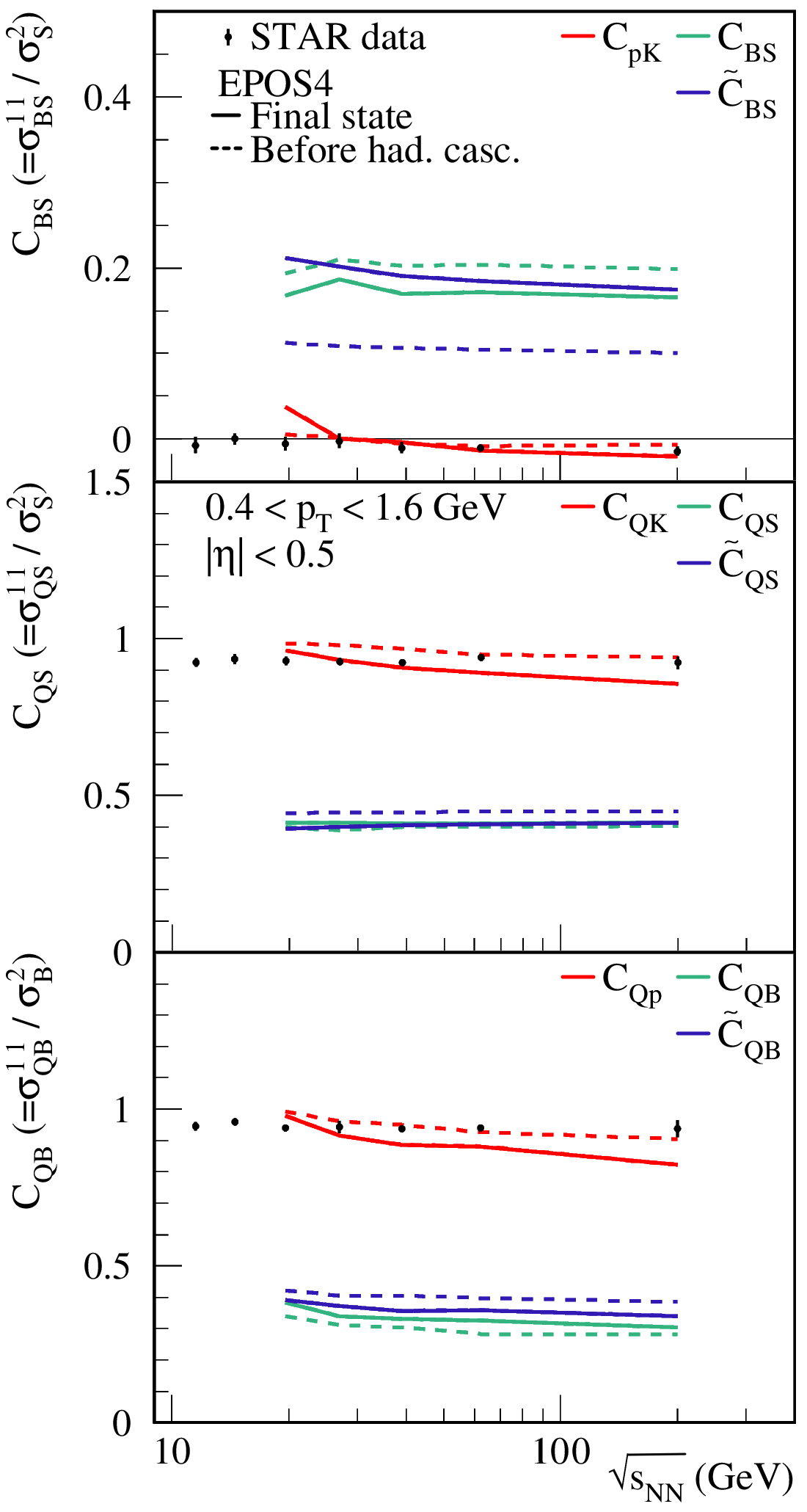}
    \caption{
    Ratios of exact $B$, $Q$ and $S$ conserved charge (co)variances and their proxies plotted as functions of $\sqrt{s_{NN}}$ for central collisions (0-5\% centrality class). Results are displayed from particle distribution in the final-state (full lines) and before hadronic cascades (dashed lines), from EPOS4 simulations. STAR data (dots) are shown as a reference 
    \cite{STAR:2019ans}.
    }
    \label{fig:ratios_engy_proxies_EPOS4_nohacas}
\end{figure}

In Fig. \ref{fig:ratios_engy_proxies_EPOS4_nohacas}, we show all the different ratios and their corresponding proxies as functions of collision energy for the most central (0-5\%) collisions, for final-state particle distributions like in Fig. \ref{fig:ratios_engy_proxies_EPOS4} (full lines), but this time compared with their associated value measured at hadronization (dashed lines). As explained in Section \ref{sec:EPOS}, the latter can be measured thanks to the modular construction of EPOS4, which allows to access particle distributions just after hadronization of the core, before re-interactions between formed hadrons are simulated (using UrQMD in cascade mode).
\\

One can see that the magnitude of almost all ratios is affected very little by the evolution of the system during hadronic cascades, as results for particle distributions at hadronization are very close to their associated value for finale-state distributions. 
While this result was expected in the case of ratios of conserved charge cumulants, it demonstrates that most proxies discussed here are somehow stable quantities, in the sense that they conserve their magnitude even after the system has gone through hadronic re-interactions which modify its chemistry.
Although we have shown in Subsection \ref{subsec:results_(co)variances} that the (co)variances used to build the proxy ratios are largely impacted by this phase in the evolution of the system, it appears that the modifications of both their respective numerator and denominator compensate overall to leave most of these ratios almost unaffected.
\\

There is one exception to this picture for the proxy ratio $\tilde{C}_{BS}$, which gets a significant modification through hadronic re-interactions. Following the same logic as in the previous paragraph, we understand this to be due to the fact that $\tilde{C}_{BS}$ is built upon $\sigma^2_\Lambda$ and $\sigma^2_K$, which are the variances that are affected respectively the most and the least from hadronic cascades.
In particular, looking at how $\tilde{C}_{BS}$ is constructed in Eq. \eqref{eq:C_BS_HRG}, we see that the numerator consist of $\sigma^2_\Lambda$ alone, which gets a growth of about $\sim4$ through the hadronic cascades, while $\sigma^2_K$ in the denominator is only increased by a factor $\sim2$. Even though the latter is added to $\sigma^2_\Lambda$, their relative difference in magnitude makes $\sigma^2_K$ dominate the denominator, hence explaining the increase of $\tilde{C}_{BS}$ after hadronic re-scatterings.
Such an important growth of net-$\Lambda$ variance, relative to other hadronic species, might come from the fact that $\Lambda$ baryons receive an important feed-down contribution from resonance decays, which are not included when considering particles at hadronization time, while they are taken into account when looking at particle distributions in the final-state.
Moreover, $\Lambda$ baryons are also less affected by baryon annihilation than protons, as discussed in  \cite{Steinheimer:2017vju}, which leads to a smaller decrease of $\sigma^2_\Lambda$ from this process as compared to $\sigma^2_p$, since their net-multiplicity will decrease less. 
\\

\begin{figure}[!h]
    \centering
    \includegraphics[width=0.8\linewidth]{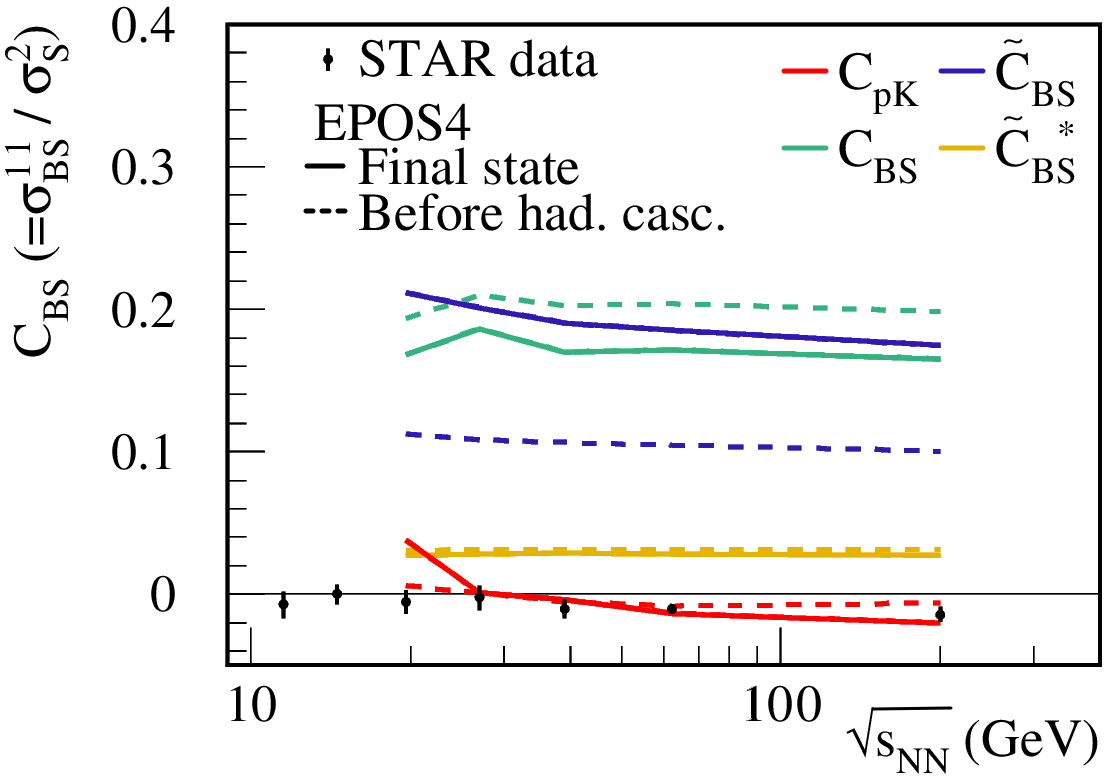}
    \caption{
    ${C}_{BS}$ ratio and different corresponding proxies plotted as functions of $\sqrt{s_{NN}}$ for central collisions (0-5\% centrality class). Results are displayed for full event calculations (full lines) and calculations before hadronic cascades (dashed lines) from EPOS4 simulations. STAR data (dots) are shown as a reference
    \cite{STAR:2019ans}.
    }
    \label{fig:ratios_C_BS_proxies_EPOS4_nohacas}
\end{figure}

In order to investigate this issue further, we tried to find another proxy ratio for $C_{BS}$ which would be less modified than $\tilde{C}_{BS}$. Thus, we have computed the following ratio:
\begin{equation}
    \Tilde{C}^*_{BS} 
    = \frac{\sigma^2_\Xi}{\sigma^2_K + 4\sigma^2_\Xi} \; ,
    \label{eq:C_BS_HRGXi}
\end{equation}
replacing variance of net-$\Lambda$ by variance of net-$\Xi$ which, even though produced less abundantly in the system, carries more strangeness and is less impacted by hadronic cascades. This quantity is compared with ${C}_{BS}$, $\tilde{C}_{BS}$ and ${C}_{pK}$ in Fig. \ref{fig:ratios_C_BS_proxies_EPOS4_nohacas}, shown in solid yellow for final state particle distributions and in dashed yellow for distributions before hadronic cascades. 
We see that although $\Tilde{C}^*_{BS}$  suffers almost no change from the hadronic re-interactions phase, it shows a large discrepancy in magnitude with $C_{BS}$, because the $\Xi$ baryons carry too small a fraction of baryon-strangeness correlation in the system. For this reason, $\Tilde{C}_{BS}$ remains the best proxy, especially since it matches $C_{BS}$ very well qualitatively in the final state, which is the only quantity accessible in experiments.

\subsection{Effective contribution from bulk matter}

\begin{figure}
    \centering
    \includegraphics[width=0.8\linewidth]{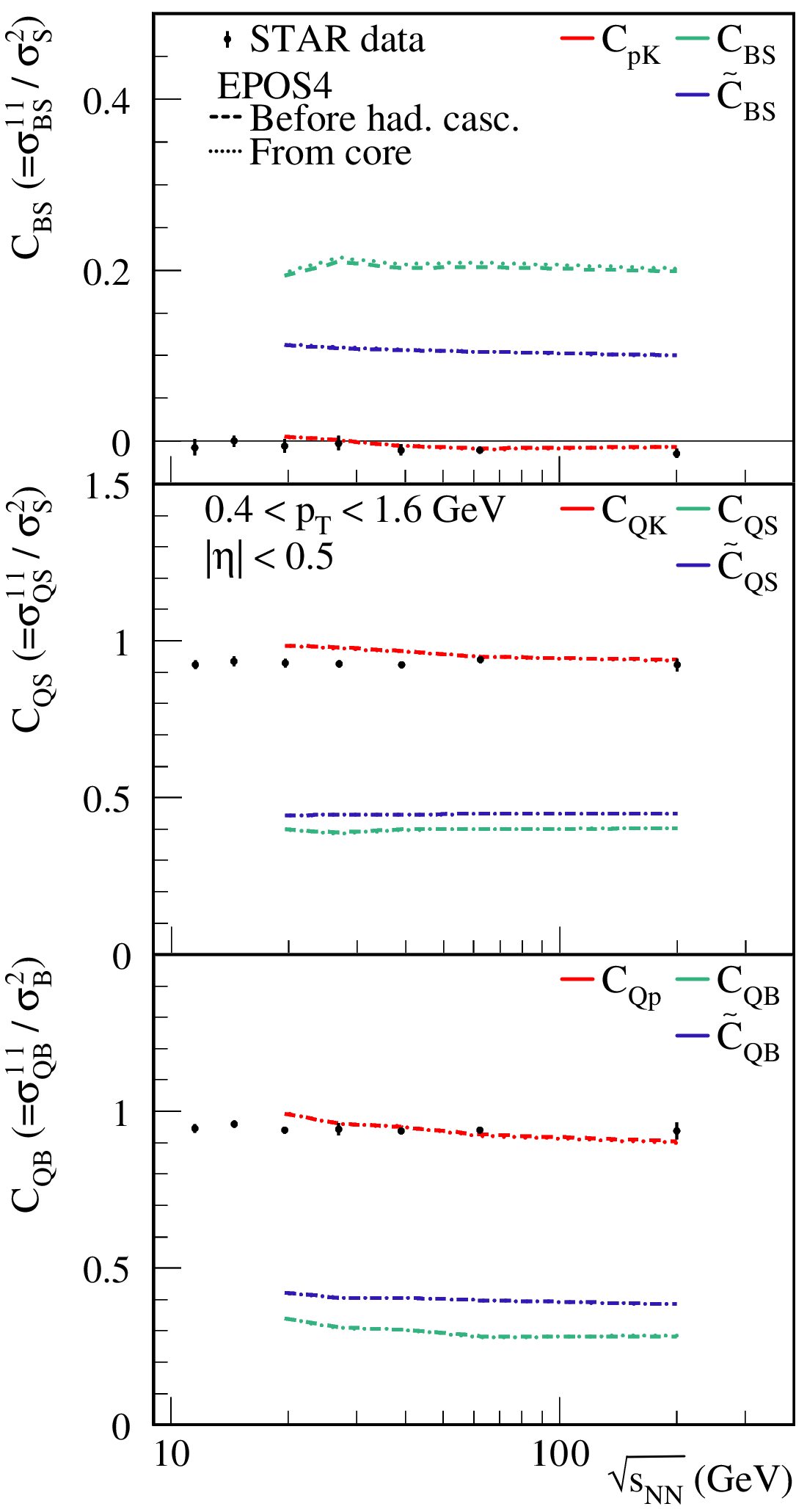}
    \caption{
    Core proportion of $C_{XY}$ and enhanced proxies $\Tilde{C}_{XY}$ ratios over the total value calculated before hadronic cascades, plotted as a function of $\sqrt{s_{NN}}$ for central collisions (0-5\% centrality class).
    Ratios of exact $B$, $Q$ and $S$ conserved charge (co)variances and their proxies plotted as functions of $\sqrt{s_{NN}}$ for central collisions (0-5\% centrality class). Results are displayed from particle distribution before hadronic cascades (dashed lines) and from core particles only (dotted lines), from EPOS4 simulations. STAR data (dots) are shown as a reference 
    \cite{STAR:2019ans}.
    }
    \label{fig:ratios_engy_proxies_EPOS4_core-over-nohacas}
\end{figure}

Finally, we show in Fig. \ref{fig:ratios_engy_proxies_EPOS4_core-over-nohacas} the different ratios from particles at hadronization time (dashed lines), compared with the same quantities computed exclusively from particles originating from the core (dotted lines), \textit{i.e.} the part of the system mimicking the behaviour of the QGP. 
Like in the previous figures, they are displayed as functions of the collision energy, for the most central (0-5\%, left panel) and most peripheral (0-80\%, right panel) collisions.
\\

We observe that for every ratio, in both central and peripheral centrality classes, the ratios computed only from core particles at hadronization are overlapping almost perfectly with the ratios computed from all particles at hadronization.
This means that the signal measured by the ratios of second order cumulants in the phase space considered here ($|\eta| < 0.5$ and $0.4 < p_T < 1.6$ GeV) is almost exclusively originating from the bulk matter, modeled by the core in EPOS4.
Hence, if the system created in such collisions would endure critical behavior, these ratios should indeed be sensitive to it, and carry its signature as long as they are measured in a similar phase space.

\section{\label{sec:Conclusion}Summary \& Outlooks}

We discussed the use of new proxies for conserved charge cumulants, which are of primary importance in the study of correlations between conserved charges $B$, $Q$ and $S$, and expected to be key observables in the search for the existence of a critical point in the phase diagram of nuclear matter. 
The new quantities $\Tilde{C}_{QS}$ and $\Tilde{C}_{BS}$, proposed in \cite{Bellwied:2019pxh}, as well as $\Tilde{C}_{QB}$ which we proposed based on that previous study,
are compared to corresponding proxies commonly measured in experiment $C_{QK}$, $C_{pK}$ and $C_{Qp}$, as well as their counterparts based on exact conserved charge quantities.
We employed EPOS4 to evaluate all these observables through simulations of Au+Au collisions at energies $\sqrt{s_{NN}} = 19.6 - 200$ GeV, where data was also available for comparison in \cite{STAR:2019ans}, in order to assess the actual performance of the new proxies in the context of a realistic out-of-equilibrium system, close to the one created experimentally.

The results of our study showed that the newly proposed proxies are better at reproducing their corresponding conserved charge cumulant ratios, as compared to the equivalent STAR proxies.
We note first of all that, with the only exception of $\Tilde{C}_{BQ}$, all these quantities are uniquely built upon variances. This makes sense, since correlations between different conserved charges are naturally carried by some hadron species (e.g. $B$ and $Q$ by the protons, $Q$ and $S$ by kaons or $B$ and $S$ by $\Lambda$ baryons).
Moreover, we observe that the addition of $\sigma^2_\Lambda$ alone to the already measured  $\sigma^2_\pi$, $\sigma^2_K$ and $\sigma^2_p$ is enough to build those new proxies. Luckily, this quantity was already measured (although in a different phase space) by STAR and published in \cite{STAR:2020ddh}. 

We compared the different ratios measured from final-state particle distributions, to their value measured from particle distribution at hadronization time. This way, we were able to quantify the effect of hadronic cascades on these quantities. 
While the ratio of conserved charge cumulants are almost not affected by hadronic re-scatterings, as expected, we can also see that $\Tilde{C}_{QB}$ and $\Tilde{C}_{QS}$ are also left unchanged after the hadronic cascades. For this reason, they represent excellent proxies of their corresponding ratios of conserved charge cumulants ${C}_{QB}$ and ${C}_{QS}$.
The only significant deviation between measurement at hadronization and in the final-state is observed for $\Tilde{C}_{BS}$. We saw that this discrepancy originates from the fact that $\sigma^2_\Lambda$ is more enhanced by hadronic cascades than other net-variances, in particular $\sigma^2_K$. Potential reasons to explain this difference have been discussed.

Finally, we investigated how much of the signal, for the phase space used in this analysis, is actually coming from particles originating from the bulk matter of the system. Unsurprisingly, the quasi-totality of the signal effectively originates from the core in EPOS4, meaning that measuring net-cumulants of hadronic species within $|\eta| < 0.5$ and with $0.4 < p_T < 1.6$ GeV actually probes fluctuations coming uniquely from bulk matter. Hence, if the system created in such collisions was to experience the impact of a critical point during its evolution, fluctuation observables should carry its imprint.
\\

It would be interesting to test the ability of the new proxy ratios $\Tilde{C}_{QB}$, $\Tilde{C}_{QS}$ and $\Tilde{C}_{BS}$ to reproduce their corresponding ratios of cumulants of conserved charges at lower collision energies, compared to the ones shown in this work. 
Most of the studies published in recent years seem to point at a collision-energy range $1-20$ GeV/A to observe phenomena which could hint at the existence of a critical point \cite{Bzdak:2019pkr}. Hence, using transport models which are suited for modeling the dynamics of HICs at these energies seems highly relevant and timely; see \cite{TMEP:2022xjg} for a recent review of the status of such models.

Investigating with more details the impact of the hadronic re-scatterings on the new proxy ratios, in particular within the previously mentioned collision-energy range, is also important to ensure their robustness through the evolution of the system. Some work in this direction has already been initiated, for instance by using an extended particle list to study the impact of resonances in \cite{Hammelmann:2023aza}, or by including mean-field potentials to model hadron interactions in a more realistic manner in transport approaches \cite{Sorensen:2020ygf}.
\\


\section*{Acknowledgments}

J.J. thanks Jean-Yves Ollitrault, Marlene Nahrgang, Grégoire Pihan and Voldymyr Vovchenko for the instructive discussions about this work.
This material is based upon work supported by the National Science Foundation under grants No. PHY-2208724,
PHY-1654219 and PHY-2116686, and within the framework
of the MUSES collaboration, under grant number No. OAC-
2103680. This material is also based upon work supported
by the U.S. Department of Energy, Office of Science, Office of Nuclear Physics, under Award Number DE-SC0022023.

\bibliography{apssamp}

\end{document}